\documentclass[10pt]{article}
\usepackage{latexsym,graphicx}
\newcommand{\be}{\begin{equation}}
\newcommand{\ee}{\end{equation}}
\def\n{\noindent}
\catcode `\@=11
\catcode `\@=12
\begin{document}
\begin{center}
\large{\bf {Bianchi Type I Massive String Magnetized Barotropic Perfect Fluid Cosmological Model in 
General Relativity}} \\
\vspace{10mm}
\normalsize{Raj Bali $^1$, Umesh Kumar Pareek$^2$ and Anirudh Pradhan$^3$}\\
\vspace{5mm}
\normalsize{$^{1}$Department of Mathematics, University of Rajasthan, Jaipur-302 004, India \\
E-mail : balir5@yahoo.co.in}\\
\vspace{5mm}
\normalsize{$^{2}$Department of Mathematics, Jaipur Engineering College and Research Centre, 
Jaipur-303 905, India \\
E-mail : ukpareek69@yahoo.co.in}\\
\vspace{5mm}
\normalsize{$^{3}$Department of Mathematics, Hindu Post-graduate College, 
Zamania-232 331, Ghazipur, India \\
E-mail : pradhan@iucaa.ernet.in, acpradhan@yahoo.com} \\
\end{center}
\vspace{10mm}
\begin{abstract} 
Bianchi type I massive string cosmological model with magnetic field of barotropic perfect 
fluid distribution through the techniques used by Latelier and Stachel, is investigated. 
To get the deterministic model of the universe, it is assumed that the universe 
is filled with barotropic perfect fluid distribution. The magnetic field is due to electric 
current produced along x-axis with infinite electrical conductivity. The behaviour of the 
model in presence and absence of magnetic field together with other physical aspects is 
further discussed. 
\end{abstract}
 \smallskip
\n Keywords: Massive string, magnetic field, Bianchi type I model, perfect fluid\\
\n PACS: 98.80.Cq, 04.20.-q 
\section{Introduction}
The cosmic strings play an important role in 
the study of the early universe. These strings arise during the phase transition after the 
big bang explosion as the temperature drops down below some critical temperature as 
predicted by grand unified theories [1-5]. It is thought that cosmic strings cause density 
perturbations leading to the formation of galaxies [6]. These cosmic strings have stress-energy 
and couple with the gravitational field. Therefore, it is interesting to study the gravitational 
effects that arise from strings. The general relativistic treatment of strings was started  by 
Letelier [7, 8] and Stachel [9]. Exact solutions of string cosmology in various space-times 
have been studied by several authors [10-23].

On the other hand, the magnetic field has an important role at the cosmological scale and is present 
in galactic and intergalactic spaces. The importance of the magnetic field for various astrophysical 
phenomena has been studied in many papers. Melvin [24] has pointed out that during the evolution of the 
universe, the matter was in a highly ionized state and is smoothly coupled with the field and forms 
a neutral matter as a result of universe expansion. FRW models are approximately valid as present 
day magnetic field strength is very small. In the early universe, the strength might have been appreciable. 
The break-down of isotropy is due to the magnetic field. Therefore the possibility of the presence of 
magnetic field in the cloud string universe is not unrealistic and has been investigated by many 
authors [25-28].

In this paper, we have investigated Bianchi type I massive string magnetized barotropic perfect 
fluid cosmological model in General Relativity. The magnetic field is due to an electric current 
produced along x-axis with infinite electrical conductivity. Also the behaviour of the model in the 
presence and absence of magnetic field together with other physical aspects is discussed. \\
\section{The Metric and Field Equations}
We consider the space-time of Bianchi type-I in the form 
\begin{equation}
\label{eq1}
ds^{2} =  - dt^{2} + A^{2}(t) dx^{2} + B^{2}(t) dy^{2} + C^{2}(t) dz^{2}.
\end{equation}
The energy momentum tensor for a cloud of massive string and perfect fluid distribution with 
electromagnetic field is taken as
\begin{equation}
\label{eq2}
T^{j}_{i} = (\rho + p) v_{i}v^{j} + p g^{j}_{i} - \lambda x_{i}x^{j} + E^{j}_{i},
\end{equation}
where $v_{i}$ and $x_{i}$ satisfy condition
\begin{equation}
\label{eq3}
v^{i} v_{i} = - x^{i} x_{i} = -1, \, \, \, v^{i} x_{i} = 0,
\end{equation}
$p$ is the isotropic pressure, $\rho$ is the proper energy density for a cloud 
string with particles attached to them, $\lambda$ is the string tension density, $v^{i}$ the 
four-velocity of the particles, and $x^{i}$ is a unit space-like vector representing 
the direction of string. In a co-moving co-ordinate system, we have
\begin{equation}
\label{eq4}
v^{i} = (0, 0, 0, 1), \, \, \, x^{i} = \left(\frac{1}{A}, 0, 0, 0 \right).
\end{equation}
The electromagnetic field $E^{j}_{i}$ given by Lichnerowicz [29] as
\begin{equation}
\label{eq5}
E^{j}_{i} = \bar{\mu}\left[\mid h \mid^{2}\left(v_{i}v^{j} + \frac{1}{2}g^{j}_{i}\right) - 
h_{i}h^{j}\right].
\end{equation}
Here the flow-vector $v_{i}$ satisfies 
\begin{equation}
\label{eq6}
g_{ij} v^{i} v^{j} = - 1, 
\end{equation}
and $\bar{\mu}$ is the magnetic permeability, $h_{i}$ the magnetic flux vector defined by 
\begin{equation}
\label{eq7}
h_{i} = \frac{\sqrt{-g}}{2\bar{\mu}}\epsilon_{ijkl}F^{kl}v^{j},
\end{equation}
where $F_{kl}$ is the electromagnetic field tensor and $\epsilon_{ijkl}$ is the Levi Civita 
tensor density. The incidental magnetic field is taken along $x$-axis, so that $h_{1} \ne 0$, 
$h_{2} = h_{3} = h_{4} = 0$. We assume that $F_{23}$ is the only non-vanishing component of $F_{ij}$.

The Maxwell's equations
\[
F_{ij;k} + F_{jk;i} + F_{ki;j} = 0,
\]
\begin{equation}
\label{eq8}
F^{ij}_{;j} = 0,
\end{equation}
are satisfied by 
$$ F_{23} =  constant = H (say). $$
Here $F_{14} = 0 = F_{24} = F_{34}$, due to the assumption of infinite electrical conductivity [30]. 
Hence 
\begin{equation}
\label{eq9}
h_{1} = \frac{AH}{\bar{\mu}BC}.
\end{equation}
Since $\mid h \mid ^{2} = h_{l}h^{l} = h_{1}h^{1} = g^{11}(h_{1})^{2}$, therefore 
\begin{equation}
\label{eq10}
\mid h \mid^{2} = \frac{H^{2}}{\bar{\mu}^{2}B^{2}C^{2}}.
\end{equation}
Using Eqs. (\ref{eq9}) and (\ref{eq10}) in (\ref{eq5}), we have 
\begin{equation}
\label{eq11}
E^{1}_{1} = - \frac{H^{2}}{2\bar{\mu}B^{2}C^{2}} = - E^{2}_{2} = - E^{3}_{3} = E^{4}_{4}. 
\end{equation}
If the particle density of the configuration is denoted by 
$\rho_{p}$, then we have
\begin{equation}
\label{eq12}
\rho = \rho_{p} + \lambda.
\end{equation}
The Einstein's field equations (in gravitational units $c = 1$, $G = 1$) read as
\begin{equation}
\label{eq13}
R^{j}_{i} - \frac{1}{2} R g^{j}_{i} = - T^{j}_{i},
\end{equation}
where $R^{j}_{i}$ is the Ricci tensor; $R$ = $g^{ij} R_{ij}$ is the
Ricci scalar. 

The field equations (\ref{eq13}) with (\ref{eq2}) subsequently lead to the following 
system of equations:
\begin{equation}
\label{eq14}
\frac{B_{44}}{B} + \frac{C_{44}}{C} + \frac{B_{4} C_{4}}{B C} = - p + \lambda + \frac{H^{2}}
{2\bar{\mu}B^{2}C^{2}},
\end{equation}
\begin{equation}
\label{eq15}
\frac{A_{44}}{A} + \frac{C_{44}}{C} + \frac{A_{4} C_{4}}{A C} = - \left[p + \frac{H^{2}}
{2\bar{\mu}B^{2}C^{2}}\right],
\end{equation}
\begin{equation}
\label{eq16}
\frac{A_{44}}{A} + \frac{B_{44}}{B} + \frac{A_{4} B_{4}}{A B} = - \left[p + \frac{H^{2}}
{2\bar{\mu}B^{2}C^{2}}\right],
\end{equation}
\begin{equation}
\label{eq17}
\frac{A_{4}B_{4}}{AB} + \frac{A_{4}C_{4}}{AC} + \frac{B_{4}C_{4}}{BC} = \rho + \frac{H^{2}}
{2\bar{\mu}B^{2}C^{2}},
\end{equation}
where the suffix $4$ at the symbols $A$, $B$ and $C$ denotes ordinary differentiation 
with respect to $t$.
\section{Solution of Field Equations}
The field Eqs. (\ref{eq14})-(\ref{eq17}) are a system of four equations with six 
unknown parameters $A$, $B$, $C$, $p$, $\lambda$ and $\rho$. Two additional constraints 
relating these parameters are required to obtain explicit solutions of the system. 

From Eq. (\ref{eq16}), we have
\begin{equation}
\label{eq18}
p = - \frac{A_{44}}{A} - \frac{B_{44}}{B} - \frac{A_{4} B_{4}}{A B} -\frac{K}{B^{2}C^{2}}, 
\end{equation}
where $ K = \frac{H^{2}}{2\bar{\mu}}$.
Now from Eq. (\ref{eq17}), we have
\begin{equation}
\label{eq19}
\rho = \frac{A_{4}B_{4}}{AB} + \frac{A_{4}C_{4}}{AC} + \frac{B_{4}C_{4}}{BC} - \frac{K}{B^{2}C^{2}}. 
\end{equation}
To get deterministic solution, we first assume that the universe is filled with barotropic perfect fluid 
which leads to
\begin{equation}
\label{eq20}
p = \gamma \rho,
\end{equation}
where $\gamma(0 \leq \gamma \leq 1)$ is a constant. Putting the values of $p$ and $\rho$ from Eqs. (\ref{eq18}) 
and (\ref{eq19}) in (\ref{eq20}), we obtain
\begin{equation}
\label{eq21}
\frac{A_{44}}{A} + \frac{B_{44}}{B} + (1 + \gamma)\frac{A_{4}B_{4}}{AB} + \gamma \frac{A_{4}C_{4}}
{AC} + \gamma \frac{B_{4}C_{4}}{BC} + (1 - \gamma)\frac{K}{B^{2}C^{2}} = 0.
\end{equation}
Equations (\ref{eq15}) and (\ref{eq16}) lead to 
\begin{equation}
\label{eq22}
\frac{(CB_{4} - BC_{4})_{4}}{(CB_{4} - BC_{4})} = - \frac{A_{4}}{A},
\end{equation}
which again leads to
\begin{equation}
\label{eq23}
C^{2}\left(\frac{B}{C}\right)_{4} = \frac{L}{A},
\end{equation}
where $L$ is an integrating constant and 
\begin{equation}
\label{eq24}
BC = \mu, ~ ~ ~ \frac{B}{C} = \nu.
\end{equation}
Thus from Eqs. (\ref{eq23}) and (\ref{eq24}), we have
\begin{equation}
\label{eq25}
\mu\left(\frac{\nu_{4}}{\nu}\right) = \frac{L}{A}.
\end{equation}
For deterministic solution, we secondly assume
\begin{equation}
\label{eq26}
A = constant = \alpha (say).
\end{equation}
Thus Eq. (\ref{eq25}) leads to
\begin{equation}
\label{eq27}
\frac{\nu_{4}}{\nu} = \frac{L}{\alpha \mu}.
\end{equation}
From Eqs. (\ref{eq21}) and (\ref{eq26}), we have
\begin{equation}
\label{eq28}
\frac{B_{44}}{B} + \gamma \frac{B_{4}C_{4}}{BC} + \frac{(1 - \gamma)K}{B^{2}C^{2}} = 0.
\end{equation}
Using (\ref{eq24}) in Eq. (\ref{eq28}), we obtain 
\begin{equation}
\label{eq29}
\frac{\mu_{44}}{2\mu} + (\gamma - 1)\frac{\mu^{2}_{4}}{4\mu^{2}} + (1 - \gamma)\frac{L^{2}}
{4\alpha^{2}\mu^{2}} + (1 - \gamma)\frac{K}{\mu^{2}} = 0,
\end{equation}
which again leads to
\begin{equation}
\label{eq30}
2\mu_{44} + (\gamma - 1)\frac{\mu^{2}_{4}}{\mu^{2}} = \frac{a}{\mu},
\end{equation}
where
\begin{equation}
\label{eq31}
a = (\gamma - 1)\frac{L^{2}}{\alpha^{2}} + 4(\gamma - 1)K.
\end{equation}
Let us assume that $\mu_{4} = f(\mu)$. Thus $\mu_{44} = ff'$, where $f' = \frac{df}{d\mu}$. 
Accordingly Eq. (\ref{eq30}) leads to
\begin{equation}
\label{eq32}
\frac{d}{d\mu}(f^{2}) + (\gamma - 1)\frac{1}{\mu}f^{2} = \frac{a}{\mu},
\end{equation}
which again reduces to
\begin{equation}
\label{eq33}
f^{2} = \left(\frac{d\mu}{dt}\right)^{2} = \frac{a}{\gamma - 1} + b\mu^{1 - \gamma}.
\end{equation}
Now from Eq. (\ref{eq27}), we have
\begin{equation}
\label{eq34}
\frac{d\nu}{\nu} = \frac{L}{\alpha \mu} \frac{dt}{d\mu} d\mu
\end{equation}
Using Eq. (\ref{eq33}) in Eq. (\ref{eq34}), we have
\begin{equation}
\label{eq35}
\frac{d\nu}{\nu} = \frac{L\bar{\mu}^{\gamma}d\mu}{\alpha \sqrt{b} \mu^{1 - \gamma} 
\sqrt{\ell^{2} + \mu^{1 - \gamma}}},
\end{equation}
where $\ell^{2} = \frac{a}{(\gamma - 1)b}$.

Eq. (\ref{eq35}), after integration, leads to 
\begin{equation}
\label{eq36}
\nu = S\left[\frac{\sqrt{\ell^{2} + \mu^{1 - \gamma}} - \ell}{\sqrt{\ell^{2} + \mu^{1 - \gamma}} 
+ \ell}\right]^{\frac{L}{\alpha(1 - \gamma)\ell \sqrt{b}}},
\end{equation}
where S is the constant of integration.

Thus the metric (\ref{eq1}) reduces to the form
\[
ds^{2} = - \left(\frac{dt}{d\mu}\right)^{2} d\mu^{2} + \alpha^{2} dx^{2} + \mu \Biggl[ S\Big[\frac{\sqrt{\ell^{2}
+ \mu^{1 - \gamma}} - \ell}{\sqrt{\ell^{2} + \mu^{1 - \gamma}} + \ell}\Big]^{\frac{L}
{\alpha \ell (1 - \gamma) \sqrt{b}}} dy^{2}
\]
\begin{equation}
\label{eq37}
+ \Big[\frac{\sqrt{\ell^{2} + \mu^{1 - \gamma}} - \ell}{\sqrt{\ell^{2} + \mu^{1 - \gamma}} 
+ \ell}\Big]^{- \frac{L}{\alpha \ell (1 - \gamma) \sqrt{b}}}dz^{2}\Biggr],
\end{equation}
which after suitable transformation of coordinates, leads to
\[
ds^{2} = - \frac{dT^{2}}{b(\ell^{2} + T^{1 - \gamma})} + dX^{2} + T \Biggl[ \Big[\frac{\sqrt{\ell^{2}
+ \mu^{1 - \gamma}} - \ell}{\sqrt{\ell^{2} + \mu^{1 - \gamma}} + \ell}\Big]^{\frac{L}
{\alpha \ell (1 - \gamma) \sqrt{b}}} dY^{2}
\]
\begin{equation}
\label{eq38}
+ \Big[\frac{\sqrt{\ell^{2} + \mu^{1 - \gamma}} - \ell}{\sqrt{\ell^{2} + \mu^{1 - \gamma}} 
+ \ell}\Big]^{- \frac{L}{\alpha \ell (1 - \gamma) \sqrt{b}}}dZ^{2}\Biggr],
\end{equation}
where $\alpha x = X, \sqrt{S}y = Y, \frac{1}{\sqrt{S}}z = z, \mu = T$.

In the absence of the magnetic field, i.e. when $K \to 0$, then the metric (\ref{eq37}) reduces to
\[
ds^{2} = - \frac{dT^{2}}{b\left(\frac{L^{2}}{\alpha^{2}b} + T^{1 - \gamma}\right)} + dX^{2} + 
T\Biggl[\Big[\frac{\sqrt{\frac{L^{2}}{\alpha^{2}b} + T^{1 - \gamma}} - \frac{L}{\alpha \sqrt{b}}}
{\sqrt{\frac{L^{2}}{\alpha^{2}b} + T^{1 - \gamma}} + \frac{L}{\alpha \sqrt{b}}}\Big]^{\frac{1}
{1 - \gamma}} dY^{2}
\]
\begin{equation}
\label{eq39}
+ \Big[\frac{\sqrt{\frac{L^{2}}{\alpha^{2}b} + T^{1 - \gamma}} - \frac{L}{\alpha \sqrt{b}}}
{\sqrt{\frac{L^{2}}{\alpha^{2}b} + T^{1 - \gamma}} + \frac{L}{\alpha \sqrt{b}}}\Big]^{-\frac{1}
{1 - \gamma}} dZ^{2}\Biggr].
\end{equation}
\section{The Geometric and Physical Significance of Model}
The energy density $(\rho)$, the string tension density $(\lambda)$, the particle density $(\rho_{p})$, 
the isotropic pressure $(p)$, the scalar of expansion $(\theta)$, and shear tensor $(\sigma)$ 
for the model (\ref{eq38}) are given by  
\begin{equation}
\label{eq40}
\rho = \frac{1}{4T^{2}}\left[b(\ell^{2} + T^{1 - \gamma}) - \frac{L^{2}}{\alpha^{2}} - 4K \right],
\end{equation}
\begin{equation}
\label{eq41}
\lambda = \frac{b(1 - \gamma)T^{1 - \gamma}}{4T^{2}} - \frac{2K}{T^{2}},
\end{equation}
\begin{equation}
\label{eq42}
\rho_{p} = \frac{1}{4T^{2}}\left[b(\ell^{2} + \gamma T^{1 - \gamma}) - \frac{L^{2}}{\alpha^{2}} + 4K \right],
\end{equation}
\begin{equation}
\label{eq43}
p = \frac{\gamma}{4T^{2}}\left[b(\ell^{2} + T^{1 - \gamma}) - \frac{L^{2}}{\alpha^{2}} 
- 4K \right].
\end{equation}
\begin{equation}
\label{eq44}
\theta = \frac{\sqrt{b}\sqrt{\ell^{2} + T^{1 - \gamma}}}{T},
\end{equation}
\begin{equation}
\label{eq45}
\sigma = \frac{1}{6\sqrt{2}T}\left[6b(\ell^{2} + T^{1 - \gamma}) + \frac{18L^{2}}{\alpha^{2}}\right]^{\frac{1}{2}}.
\end{equation}
Thus
\begin{equation}
\label{eq46}
\rho + p = \frac{1}{4T^{2}}\left[b\{2\ell^{2} - (1 - \gamma)T^{1 - \gamma}\} - \frac{2L^{2}}
{\alpha^{2}} - 8K \right],
\end{equation}
and
\begin{equation}
\label{eq47}
\rho + 3p = \frac{1}{4T^{2}}\left[b\{4\ell^{2} + (1 + 3\gamma)T^{1 - \gamma} - \frac{4L^{2}}
{\alpha^{2}} - 16K \right].
\end{equation}
The reality conditions given by Ellis [31] as
$$ (i) \rho + p > 0, ~ ~ ~ (ii) \rho + 3p > 0, $$
are satisfied when $$ T^{1 - \gamma} < \frac{2}{1 - \gamma}\left[\ell^{2} - \frac{L^{2}}{b\alpha^{2}} 
- \frac{4K}{b}\right].$$

The energy conditions $\rho \geq 0$ and $\rho_{p} \geq 0$ are satisfied in the presence of magnetic 
field for the model (\ref{eq38}). The condition $\rho \geq 0$ leads to
$$ b(\ell^{2} + T^{1 - \gamma}) \geq \frac{L^{2}}{\alpha^{2}} + 4K.$$
The condition $\rho_{p} \geq 0$ leads to
$$ b(\ell^{2} + T^{1 - \gamma}) \geq \frac{L^{2}}{\alpha^{2}} - 4K.$$
From Eq. (\ref{eq42}), we observe that the string tension density $\lambda \geq 0$ provided
$$ b(1 - \gamma)T^{1 - \gamma} \geq 8K. $$ 

The model (\ref{eq38}) starts with a big bang at $T = 0$ and the expansion in the model decreases as 
time increases. When $T \to 0$ then $\rho \to \infty$, $\lambda \to \infty$. When $T \to \infty$ 
then $\rho \to 0$, $\lambda \to 0$. Also $p \to \infty$ when $T \to 0 $ and $p \to 0$ when $T \to \infty$. 
Since $ \lim_{T \to \infty}  \frac{\sigma}{\theta} \ne 0$, hence the model does not isotropize in general. 
However, if $L = 0$ then the model (\ref{eq38}) isotropizes for large values of $T$. There is a point type 
singularity [32] in the model (\ref{eq38}) at $T = 0$. \\
The ratio of magnetic energy to material energy is given by
\begin{equation}
\label{eq48}
\frac{E^{4}_{4}}{\rho} = \frac{4K}{b(\ell^{2} + T^{(1 - \gamma)}) -
  \frac{L^{2}}{\alpha^{2}} - 4K},
\end{equation}
where $0 \leq \gamma \leq 1$. The ratio $\frac{E^{4}_{4}}{\rho}$ is non-zero
finite quantity initially and tends to zero as $T \to \infty$. \\
The scale factor $(R)$ is given by
\begin{equation}
\label{eq49}
R^{3} = ABC = \alpha \mu = \alpha T.
\end{equation}
Thus $R$ increases as $T$ increases. \\
The deceleration parameter $(q)$ in presence of magnetic field is given by
\begin{equation}
\label{eq50}
q = - \frac{R R_{44}}{R^{2}_{4}} = - \frac{\left[a(3a - 2) + \frac{1}{2}b(1 + 3\gamma)T^{(1 -
      \gamma)}\right]}{\left(\frac{a}{\gamma - 1} + b T^{(1 - \gamma)}\right)}.
\end{equation}
The deceleration parameter $(q)$ approaches the value $(-1)$ as in the case of
de-Sitter universe if
$$ T^{(1 - \gamma)} = \frac{2a(3a - 2)(\gamma - 1) - a}{b(1 - \gamma)(1 - 3\gamma)}.$$
In the absence of magnetic field, i.e. $K \to 0$, the above mentioned quantities are given by
\begin{equation}
\label{eq51}
\rho = \frac{b}{4T^{1 + \gamma}},
\end{equation}
\begin{equation}
\label{eq52}
\lambda = \frac{b(1 - \gamma)}{4T^{1 + \gamma}},
\end{equation}
\begin{equation}
\label{eq53}
\rho_{p} = \frac{b \gamma}{4T^{1 + \gamma}},
\end{equation}
\begin{equation}
\label{eq54}
p = \frac{b \gamma}{4T^{1 + \gamma}},
\end{equation}
In the absence of magnetic field when $\gamma = 1$, then $\rho = \frac{b}{4T^{2}}$ and also the string 
tension density becomes zero. The energy conditions $\rho \geq 0$ and $\rho_{p} \geq 0$ are satisfied 
for the model (\ref{eq38}) when $b \geq 0$. 

The reality conditions given by Ellis [31] as
$$ (i) ~ ~ ~ \rho + p > 0, ~ ~ ~ (ii) ~ ~ ~ \rho + 3p > 0, $$
are satisfied when $b > 0$. 
\begin{equation}
\label{eq55}
\theta = \frac{\sqrt{\frac{L^{2}}{\alpha^{2}} + bT^{1 - \gamma}}}{T},
\end{equation}
\begin{equation}
\label{eq56}
\sigma = \frac{1}{6\sqrt{2}T}\left[\frac{24L^{2}}{\alpha^{2}} + 6b T^{1 - \gamma}\right]^{\frac{1}{2}}.
\end{equation}
In the absence of magnetic field, the model (\ref{eq39}) starts with a big bang at $T = 0$ and the expansion 
in the model decreases as time increases. When $T \to 0$ then $\rho \to \infty$, $\lambda \to \infty$ 
and $p \to \infty$. When $T \to \infty$ then $\rho \to 0$, $\lambda \to 0$ and $p \to 0$. In the absence of 
magnetic field, the particle density $(\rho_{p})$ and the isotropic pressure $(p)$ are equal. 
Since $ \lim_{T \to \infty}  \frac{\sigma}{\theta} \ne 0$, therefore the model does not isotropize in general. 
However, if $L = 0$ then the model (\ref{eq39}) isotropizes for large values
of $T$. There is a point type 
singularity [32] in the model (\ref{eq39}) at $T = 0$. \\
In absence of magnetic field, the scale factor $(R)$ is given by
\begin{equation}
\label{eq57}
R^{3} = \alpha T.
\end{equation}
The $R$ increases as T increases in this case also. The
deceleration parameter $(q)$ is given by
\begin{equation}
\label{eq58}
q = - \left[\frac{b(1 + 3\gamma)T^{(1 - \gamma)} -
  \frac{L^{2}}{\alpha^{4}}\{3L^{2}(1 - \gamma) +
  \alpha^{2}\}}{\left(\frac{L^{2}}{\alpha^{2}} + bT^{(1 - \gamma)}\right)}\right]
\end{equation}  
We observe that $q < 0$ if $T^{(1 - \gamma)} > \frac{L^{2}\{3L^{2}(1 -
  \gamma)\alpha^{2}\}}{b(1 + 3\gamma)\alpha^{4}}$. 
The deceleration parameter $(q)$ approaches the value $(-1)$ as in the case of
de-Sitter universe if
$$ T^{(1 - \gamma)} = \frac{L^{2}\{3L^{2}(1 -
  \gamma) + \alpha^{2}\}}{3b\gamma \alpha^{4}} $$
\section*{Acknowledgments} 
Authors would like to thank the Inter-University Centre for Astronomy and Astrophysics
(IUCAA), Pune, India for providing facility and support where this work was
carried out. Authors also thank to the referee for their fruitful comments.  

\end{document}